\documentclass[aps,prd,showpacs,preprintnumbers,twocolumn,nofootinbib]{revtex4}
\usepackage[dvips]{graphicx}
\usepackage{bm,latexsym,amsmath,amssymb,amsfonts}
\begin{document}

\title{
Low energy effective gravitational equations on a Gauss-Bonnet brane
}

\author{Tsutomu~Kobayashi$^1$, Tetsuya~Shiromizu$^1$, and Nathalie~Deruelle$^2$}

\affiliation{
$^1$Department of Physics, Tokyo Institute of Technology, Tokyo 152-8551, Japan\\
$^2$Institut des Hautes Etudes Scientifiques, 35 Route de Chartres, 91440, Bures-sur-Yvette, France
}

\begin{abstract}
We present effective gravitational equations at low energies in
a $Z_2$-symmetric braneworld with the Gauss-Bonnet term.
Our derivation is based on the geometrical projection approach,
and we solve iteratively the bulk geometry using the gradient expansion scheme.
Although the original field equations are quite complicated due to
the presence of the Gauss-Bonnet term,
our final result
clearly has the form of the Einstein equations plus correction terms, which
is simple enough to handle.
As an application, we consider homogeneous and isotropic cosmology on the brane.
We also comment on the holographic interpretation of bulk gravity
in the Gauss-Bonnet braneworld. 
\end{abstract}

\pacs{04.50.+h, 98.80.Cq}

\maketitle

\section{Introduction}

Recent progress in string theory
provides a revolutionary picture of the universe,
in which our four-dimensional world
is in fact a ``braneworld''
described by the motion of a thin wall
embedded in a higher dimensional spacetime.
Various aspects on braneworld models
have been addressed so far~\cite{Maa_rev}, and
in a number of related publications
the simplest Randall-Sundrum model~\cite{RS}
is employed for the analysis of gravity on the brane.
In the Randall-Sundrum model,
five-dimensional vacuum Einstein gravity with a negative cosmological constant
is assumed.
However, there will be higher derivative corrections in string gravity
and hence it is natural to
consider braneworld models with such corrections in the bulk action.
While higher derivative corrections generally induce unwanted ill behavior,
the special combination of curvature tensors
called the {\em Gauss-Bonnet} term is known to avoid pathology:
the Lagrangian is ghost-free,
leading to well controlled field equations both at the
classical and quantum levels~\cite{Nat_rev}.
In this paper we consider on the braneworld model
described by the Einstein-Hilbert action plus
the Gauss-Bonnet term.

In the context of the Gauss-Bonnet braneworld,
cosmology~\cite{Charm, Cos_others} and linearized gravity~\cite{DS, others}
has been discussed so far in the literature.
In the present paper we consider nonlinear gravity in the braneworld.
In this direction, Maeda and Torii~\cite{MT} have
derived the effective gravitational equations on the brane using the geometrical 
projection approach.
Though in the Randall-Sundrum model
the geometrical projection is a  highly successful approach
to clarify the nature of brane gravity~\cite{SMS},
Maeda and Torii's effective equations in the Gauss-Bonnet braneworld
are too complicated to be physically transparent.
This is because the extrinsic curvature on the brane cannot be
expressed solely by the energy-momentum tensor on it
in a formal way.
To cope with this problem, 
we shall resort to an approximation method.
The approximation employed here is
that
the relevant length scales on the brane are much larger than the bulk curvature scale,
focusing on gravitational physics at low energies.

The rest of the paper is organized as follows. In the next section
we describe the model and present the basic equations 
following Ref.~\cite{MT}.
In Sec.~III, we solve the bulk equation
using the gradient expansion scheme and derive the effective equations 
on the Gauss-Bonnet brane.
As the simplest example, we discuss homogeneous and isotropic
cosmology on the brane in Sec.~IV.
Section V is devoted to summary and discussion.

\section{Basic equations}

The action we consider is given by
%
\begin{eqnarray}
S & = &  \frac{1}{2\kappa^2} \int d^5x {\sqrt {-g}}
\left({\cal R}-2\Lambda +\alpha {\cal L}_{{\rm GB}} \right) \nonumber \\
& &\quad +\int d^4 x {\sqrt {-h}}\left(Q+{\cal L}_m-\sigma\right),
\end{eqnarray}
%
where
${\cal R}$ is the Ricci scalar of the five-dimensional metric $g_{MN}$ and
$\Lambda$ is the cosmological constant in the bulk.
$h_{\mu\nu}$ is the induced metric on the brane,
$\sigma$ is the brane tension, and
${\cal L}_m$ is the Lagrangian for matter localized on the brane.
The Gauss-Bonnet term ${\cal L}_{{\rm GB}}$ is defined as
%
\begin{eqnarray}
{\cal L}_{{\rm GB}} =  {\cal R}^2
-4{\cal R}_{MN}{\cal R}^{MN} 
+{\cal R}_{MNJK}{\cal R}^{MNJK}, 
\end{eqnarray}
%
and the parameter $\alpha$ has dimension of (length)$^2$.
The surface term $Q$ is the Gibbons-Hawking-Myers term defined by \cite{BT}
%
\begin{eqnarray}
Q=2K + 4\alpha \left(J-2 G_\mu^{~\nu} K_\nu^{~\mu}\right),
\end{eqnarray}
%
where $K_\mu^{~\nu}$ is the extrinsic curvature of the brane
and $K$ is its trace,
$G_\mu^{~\nu}$ is the four dimensional Einstein tensor 
with respect to the induced metric $h_{\mu\nu}$, and $J$ is the trace of
%
\begin{eqnarray}
J_{\mu}^{~\nu} & = & 
\frac{1}{3}\left(2KK_{\mu}^{~\alpha}K_{\alpha}^{~\nu}
+K_{\alpha}^{~\beta} K_{\beta}^{~\alpha} K_{\mu}^{~\nu}\right.
\nonumber \\
& &\quad\left. -2K_{\mu}^{~\alpha}K_{\alpha}^{~\beta}K_{\beta}^{~\nu}-K^2 K_{\mu}^{~\nu}\right). 
\end{eqnarray}
%

In this paper
we shall derive effective gravitational equations on a brane
by solving the bulk geometry.
For this purpose,
we use the (4$+$1)-decomposition of the metric and
write the five-dimensional field equations in the form of
the evolution equation along the 
extra dimension and the constraint equations.
In deriving our basic equations, we will closely follow
the geometrical projection approach
by Maeda and Torii (henceforth MT)~\cite{MT}.
(See Appendix for an alternative approach based on the work by
Deruelle and Katz~\cite{DK}.)

We write the bulk metric in the form
%
\begin{eqnarray}
g_{MN}dx^M dx^N = dy^2+q_{\mu\nu}(y,x) dx^\mu dx^\nu, 
\end{eqnarray}
%
where $y$ is the fifth coordinate.
We may assume that the position of the brane is $y=0$ without loss of generality,
so that
the induced metric on the brane is $h_{\mu\nu} (x)=q_{\mu\nu}(y=0,x)$. 
We also assume a $Z_2$-symmetry across the brane.

The traceless part 
of the evolution equation is given by
%
\begin{eqnarray}
\mbox \pounds_n \tilde K_{\mu}^{~\nu}
&=&-\frac{3}{2}E_{\mu}^{~\nu}-\frac{1}{2} \tilde M_{\mu}^{~\nu} 
\nonumber\\
&&\quad-\left(
K_{\mu}^{~\alpha} K_{\alpha}^{~\nu} -\frac{1}{4}\delta_{\mu}^{~\nu}
K_{\alpha}^{~\beta} K_{\beta}^{~\alpha}
\right), \label{tlee}
\end{eqnarray}
%
where $\tilde K^\mu_\nu$ is the traceless part of the extrinsic curvature 
$K_{\mu}^{~\nu} (y, x)= (1/2)q^{\alpha\nu}\partial_y q_{\mu\alpha}$, 
\begin{eqnarray}
E_{\mu\nu}:={}^{(5)}C_{\mu \alpha \nu \beta} n^{\alpha} n^{\beta}
={}^{(5)}C_{\mu y \nu y}
\end{eqnarray}
is the ``electric'' part of the five-dimensional Weyl tensor,
and $n^{\alpha}$ is a hypersurface normal, $n^\alpha = 
(\partial_y)^\alpha$. 
$\tilde M_{\mu}^{~\nu}$ is the traceless part of
$M_{\mu}^{~\nu} := M_{\mu\alpha}^{~~~\nu\alpha}$ where
\begin{eqnarray}
M_{\mu \alpha \nu \beta}:=R_{\mu\alpha \nu \beta}[q]
-K_{\mu \nu} K_{\alpha\beta}+K_{\mu\beta}K_{\nu \alpha}.
\end{eqnarray}
and $R_{\mu\alpha \nu \beta}[q]$ is the four-dimensional Riemann tensor
of the metric $q_{\mu\nu}(y, x)$.

The traceless part of the MT effective equations on the brane is arranged as 
%
\begin{eqnarray}
& & \frac{3}{2}\left(\tilde M_{\mu}^{~\nu} + E_{\mu}^{~\nu} \right)
+\alpha \left[ {}_{(a)}\bar H_{\mu}^{~\nu} +{}_{(b)}\bar H_{\mu}^{~\nu} + 
{}_{(c)}\bar H_{\mu}^{~\nu}\right]
\nonumber \\
& & \qquad= -\frac{4\alpha}{3+\alpha M} \tilde M_{\mu}^{~\nu} \Lambda,
\end{eqnarray}
%
where $M$ is the trace of $M_{\mu}^{~\nu}$.
In the above we defined
%
\begin{eqnarray}
{}_{(a)}\bar H_{\mu\nu} & := & 
2\left(
L_{\mu \alpha\beta \gamma}L_\nu^{~\alpha\beta\gamma}
-\tilde M^{\alpha\beta}L_{\mu\alpha\nu\beta}-\tilde M_{\mu}^{~\alpha} \tilde 
M_{\nu\alpha}
\right) \nonumber \\
& & -\frac{3-\alpha M}{6(3+\alpha M)} M \tilde M_{\mu\nu} 
+\frac{2\alpha}{3+\alpha M} \tilde M_{\alpha}^{~\beta} \tilde M_{\beta}^{~\alpha}\tilde M_{\mu\nu}
\nonumber \\
& & -\frac{1}{2}q_{\mu\nu}\left(L_{\alpha\beta\rho\sigma}L^{\alpha\beta\rho\sigma}
-\tilde M_{\alpha\beta} \tilde M^{\alpha\beta}\right),
\\
{}_{(b)}\bar H_{\mu\nu} &: = & -3\left(
 \tilde M_{\mu\alpha}E^{\alpha}_{~\nu}
+\tilde M_{\nu\alpha} E^{\alpha}_{~\mu}+2L_{\mu\alpha\nu\beta}E^{\alpha\beta}
\right) \nonumber \\
& & +\frac{3}{2}q_{\mu\nu}\tilde M_{\alpha\beta}E^{\alpha\beta}
+\frac{1}{2}ME_{\mu\nu} \nonumber \\
& & +\frac{6\alpha}{3+\alpha M} \tilde M_{\alpha\beta}
E^{\alpha\beta} \tilde M_{\mu\nu},
\end{eqnarray}
%
and
%
\begin{eqnarray}
 {}_{(c)}\bar H_{\mu\nu} & := & -4N_\mu N_\nu +4N^\alpha 
 \left(N_{\alpha \mu\nu}+N_{\alpha \nu\mu}\right)
\nonumber \\&&\hspace{-5mm}
+2N_{\alpha\beta\mu}N^{\alpha\beta}_{~~~\nu}
+4N_{\mu\alpha\beta}N_\nu^{~\alpha\beta}
\nonumber\\&&\hspace{-10mm}
+3q_{\mu\nu} \left(N_\alpha N^\alpha -\frac{1}{2}N_{\alpha\beta\gamma}
N^{\alpha\beta\gamma} \right)
\nonumber \\& &\hspace{-15mm}
+\frac{4\alpha}{3+\alpha M} \left(N_\alpha N^\alpha -\frac{1}{2}N_{\alpha\beta\gamma}
N^{\alpha\beta\gamma}  \right)\tilde M_{\mu\nu},
\end{eqnarray}
%
with
%
\begin{eqnarray}
L_{\mu\nu\alpha\beta}
& := & M_{\mu\nu\alpha\beta}+q_{\mu[\beta}\tilde M_{\alpha] \nu}
+q_{\nu [\alpha}\tilde M_{\beta] \mu} \nonumber \\
& & -\frac{1}{6}M q_{\mu [\alpha} q_{\beta] \nu},
\\
N_{\mu\nu\alpha}&:=&D_\mu K_{\nu\alpha}-D_\nu K_{\mu\alpha},
\\
N_{\mu}&:=&q^{\alpha\beta}N_{\alpha \mu \beta}=D_\nu K^\nu_\mu-D_\mu K,
\end{eqnarray}
%
and $D_\mu$ is the covariant derivative with respect to
the four-dimensional metric $q_{\mu\nu}$. 

The trace part of the MT effective equations on the brane
leads to the Hamiltonian constraint:
%
\begin{eqnarray}
\frac{1}{2}M+\alpha \left[{}_{(a)}H+{}_{(b)}H+{}_{(c)}H  \right]=\Lambda,
\end{eqnarray}
%
where $M:=M_{\mu}^{~\mu}=R[q]-K^2+K_{\mu}^{~\nu}K_{\nu}^{~\mu}$.
$_{(a)}H$, $_{(b)}H$, and $_{(c)}H$ are
the trace of
%
\begin{eqnarray}
{}_{(a)}H_{\mu\nu}
& := & 2M_{\mu\alpha\beta\gamma}M_\nu^{~\alpha\beta\gamma}
-6M^{\alpha\beta}M_{\mu\alpha\nu\beta}
+4MM_{\mu\nu}
\nonumber \\
& & -8M_{\mu\alpha}M_{\nu}^{~\alpha}
-\frac{1}{8}q_{\mu\nu}
\left(7M^2-24M_{\alpha\beta}M^{\alpha\beta}\right.
\nonumber \\
& &\left. +3M_{\alpha\beta\rho\sigma}
M^{\alpha\beta\rho\sigma}\right),
\\
{}_{(b)}H_{\mu\nu}
& := & -6\left(M_{\mu\alpha}E^\alpha_{~\nu}+M_{\nu\alpha}E^\alpha_{~\mu}
+M_{\mu\alpha\nu\beta}E^{\alpha\beta}\right)
\nonumber \\
& & +\frac{9}{2}q_{\mu\nu}M_{\alpha\beta}E^{\alpha\beta}
+3ME_{\mu\nu},
\end{eqnarray}
and
\begin{eqnarray}
{}_{(c)}H_{\mu\nu}
& := & -4N_\mu N_\nu +4N^\alpha
\left(N_{\alpha \mu\nu}+N_{\alpha \nu \mu} \right) \nonumber \\
& & +2N_{\alpha\beta\mu}N^{\alpha\beta}_{~~~\nu}
+4N_{\mu\alpha\beta}N_{\nu}^{~\alpha\beta} \nonumber \\
& & +3q_{\mu\nu}
\left(N_\alpha N^\alpha -\frac{1}{2}N_{\alpha\beta\gamma}N^{\alpha\beta\gamma}  
\right). 
\end{eqnarray}

We require that an exact anti-de Sitter bulk
with the curvature radius $\ell$ solves
the five-dimensional field equations.
Then the Hamiltonian constraint gives rise to the relation
\begin{eqnarray}
\Lambda = -\frac{6}{\ell^2}\Bigl( 1-\frac{\beta}{2} \Bigr). 
\end{eqnarray}
where we introduced a useful dimensionless quantity
\begin{eqnarray}
\beta:=\frac{4\alpha}{\ell^2}. 
\end{eqnarray}


The Codacci equation, or the momentum constraint, is given by 
\begin{eqnarray}
&&D_\nu \left[K_{\mu}^{~\nu}- \delta_{\mu}^{~\nu}K\right.
\nonumber\\
&&\qquad\left.
+2\alpha 
\left(3J_{\mu}^{~\nu}-J \delta_{\mu}^{~\nu} -2P_{\mu\alpha}^{~~~\nu\beta}K_{\beta}^{~\alpha}
\right) \right]
=0,
\end{eqnarray}
where
\begin{eqnarray}
P_{\mu\nu\alpha\beta}
& = & R_{\mu\nu\alpha\beta}+2 h_{\mu [\beta} R_{\alpha] \nu}
+2h_{\nu[\alpha}R_{\beta]\mu} \nonumber \\
& &\qquad\qquad+ R h_{\mu[\alpha}h_{\beta]\nu}.
\label{def=P}
\end{eqnarray}

The (generalized) Israel junction equations give the boundary condition at the brane as 
\cite{Israel_Junk, Davis:2002gn}
%
\begin{eqnarray}
& & K_{\mu}^{~\nu}
+\frac{\beta \ell^2}{3}
\left(\frac{9}{2}J_{\mu}^{~\nu}-J\delta_{\mu}^{~\nu}  -3P_{\mu\alpha}^{~~~\nu \beta}
K_{\beta}^{~\alpha}- G_{\alpha\beta}K^{\alpha\beta}\delta_{\mu}^{~\nu}\right)
\nonumber\\
&&
=-\frac{\kappa^2}{6}\sigma \delta_{\mu}^{~\nu} 
-\frac{\kappa^2}{2}\left(T_{\mu}^{~\nu}-\frac{1}{3}\delta_{\mu}^{~\nu} T \right), \label{jun}
\end{eqnarray}
%
where $T_{\mu\nu}$ is the energy-momentum tensor on the brane.

As a remark, we note that our basic equation~(\ref{tlee}) 
holds on any $y=$constant hypersurfaces because 
the junction condition is not used in the derivation.

\section{Effective equations on the brane}

We will use the gradient expansion scheme~\cite{KS, GE} in order to solve the bulk 
evolution equations, and then will derive the four-dimensional effective equations 
describing the brane geometry. The small expansion parameter here is the 
ratio of the bulk curvature scale 
to the brane intrinsic curvature scale,
%
\begin{eqnarray}
\epsilon = \ell^2 \left|R\right|.
\end{eqnarray}
%
For any tensor $A_\mu^{~\nu}$ we expand it as 
%
\begin{eqnarray}
A_\mu^{~\nu} = {}^{(0)}A_\mu^{~\nu} + {}^{(1)}A_\mu^{~\nu} + {}^{(2)}A_\mu^{~\nu} + \cdots,
\end{eqnarray}
where $^{(n)}A_\mu^{~\nu}={\cal O}(\epsilon^n)$.

%
%
%

\subsection{Zeroth order equations}

The zeroth order solution is given by 
\begin{eqnarray}
{}^{(0)}K_{\mu}^{~\nu}  =-\frac{1}{\ell}\delta_{\mu}^{~\nu}.
\label{zerothsol}
\end{eqnarray}
Since we have
\begin{eqnarray}
{}^{(0)}M_{\mu \nu}^{~~~\alpha \beta}&=&\frac{1}{\ell^2}\left(
- \delta_\mu^{~\alpha}\delta_\nu^{~\beta} +
\delta_\mu^{~\beta} \delta_\nu^{~\alpha}\right),
\\
{}^{(0)}L_{\mu\nu}^{~~~\alpha\beta}&=&0,
\end{eqnarray}
one can confirm that Eq.~(\ref{zerothsol}) trivially solves
the evolution equation, the traceless part of the MT effective equations,
and the Codacci equation at zeroth order.
It is easy to check that Eq.~(\ref{zerothsol}) satisfies the Hamiltonian constraint as well.
The bulk metric is written in the form 
\begin{eqnarray}
{}^{(0)}g_{MN}dx^Mdx^N=dy^2+a^2(y)h_{\mu\nu}(x)dx^\mu dx^\nu,
\end{eqnarray}
where the warp factor is given by
\begin{eqnarray}
a=e^{-y/\ell}. 
\end{eqnarray}

From the junction condition we obtain the following relation
among the parameters:
%
\begin{eqnarray}
\frac{1}{\ell}\left(1-\frac{\beta}{3} \right)=\frac{1}{6}\kappa^2 \sigma. 
\end{eqnarray}
%

\subsection{First order equations}

The zeroth order result in the previous subsection is rather trivial;
going to higher order we will obtain
the field equations that govern the brane geometry $h_{\mu\nu}(x)$.

First let us look at the trace part of the extrinsic curvature,
which can be determined without solving the bulk geometry.
It is easy to see that
%
\begin{equation}
{}^{(1)}_{(a)}H=-\frac{2}{\ell^2}{}^{(1)}M,
\quad{}^{(1)}_{(b)}H={}^{(1)}_{(c)}H=0,
\end{equation}
%
and then the Hamiltonian constraint reduces to a simple
form\footnote{We assume $\beta\neq1$ throughout the paper.}:
%
\begin{equation}
(1-\beta){}^{(1)}M=0.
\end{equation}
%
Thus we arrive at the same conclusion as the
Randall-Sundrum braneworld:
\begin{equation}
{}^{(1)}K=-\frac{\ell}{6}R[q]=-\frac{\ell}{6a^2}R[h] \label{traceK}.
\end{equation}

Let us move on to the investigation of the traceless part.
It is straightforward to show that
\begin{eqnarray}
{}^{(1)}_{(a)}\bar H_{\mu}^{~\nu} & = & -\frac{3-\alpha {}^{(0)}M}{6(3+\alpha {}^{(0)}M)} {}^{(0)}M 
{}^{(1)}\tilde M_{\mu}^{~\nu} \nonumber \\
& = & \frac{2}{\ell^2}\frac{1+\beta}{1-\beta} {}^{(1)}\tilde M_{\mu}^{~\nu},
\\
{}^{(1)}_{(b)}\bar H_{\mu}^{~\nu}&=&\frac{1}{2}{}^{(0)}M{}^{(1)}E_{\mu}^{~\nu}
=-\frac{6}{\ell^2}
{}^{(1)}E_{\mu}^{~\nu} 
\\
{}^{(1)}_{(c)}\bar H_{\mu}^{~\nu} &=&0, 
\end{eqnarray}
and
\begin{eqnarray}
{}^{(1)}M_{\mu}^{~\nu}= R_{\mu}^{~\nu}[q]+\frac{2}{\ell}{}^{(1)}K_{\mu}^{~\nu} 
+\frac{1}{\ell}{}^{(1)}K \delta_{\mu}^{~\nu}.
\end{eqnarray}
Then the traceless part of the MT effective equations reads
%
\begin{eqnarray}
^{(1)}E_{\mu}^{~\nu} = -{}^{(1)}\tilde M_{\mu}^{~\nu}.
\end{eqnarray}
%
Substituting this into the traceless part of the 
evolution equation, we obtain 
%
\begin{equation}
\mbox \pounds_n {}^{(1)}\tilde K_{\mu}^{~\nu}
 =\tilde R_{\mu}^{~\nu}[q]+\frac{4}{\ell} {}^{(1)}\tilde K_{\mu}^{~\nu}, 
\label{1st_Ev}
\end{equation}
%
where ${}\tilde R_{\mu}^{~\nu}[q]$ is the traceless part of 
the Ricci tensor $R_{\mu}^{~\nu}[q]=R_{\mu}^{~\nu}[h]/a^2(y)$. 
Note that Eq.~(\ref{1st_Ev}) is again exactly the same as the
first order evolution equation in the Randall-Sundrum model.
The above equation can be integrated to give
%
\begin{equation}
{}^{(1)}\tilde K_{\mu}^{~\nu} (y, x)=-\frac{\ell}{2a^2}{}\tilde R_{\mu}^{~\nu}[h] 
+\frac{1}{a^4} {}^{(1)}\chi_{\mu}^{~\nu} (x), \label{tK}
\end{equation}
%
where ${}^{(1)}\chi_{\mu}^{~\nu} (x)$ is an integration constant
whose trace vanishes: $\chi_{\mu}^{~\mu}=0$.
To avoid confusion, we explicitly show here that the Ricci tensor in the equation
is constructed from the induced metric $h_{\mu\nu}(x)$.
Combining Eq.~(\ref{tK}) and the trace part~(\ref{traceK}), we obtain 
%
\begin{equation}
{}^{(1)}K_{\mu}^{~\nu}=-\frac{\ell}{2}
\frac{1}{a^{2}}\left(R_{\mu}^{~\nu} -\frac{1}{6} \delta_{\mu}^{~\nu} R  \right)
+\frac{1}{a^{4}}{}^{(1)}\chi_{\mu}^{~\nu}. \label{1stsol}
\end{equation}

The extrinsic curvature~(\ref{1stsol}) is related to the energy-momentum tensor
on the brane via the junction condition.
The junction condition at first order is 
%
\begin{eqnarray}
& & {}^{(1)}K_{\mu}^{~\nu}
+\frac{\beta \ell^2}{3} \left( 
\frac{9}{2}{}^{(1)}J_{\mu}^{~\nu} - {}^{(1)}J \delta_{\mu}^{~\nu} 
+\frac{3}{\ell}{}^{(1)}P_{\mu}^{~\nu} -\frac{1}{\ell}{}^{(1)}P \delta_{\mu}^{~\nu}
\right) \nonumber \\
& & \quad= -\frac{\kappa^2}{2}
\left(T_{\mu}^{~\nu} -\frac{1}{3}\delta_{\mu}^{~\nu} T \right).
\label{jnk1st}
\end{eqnarray}
%
Substituting
%
\begin{eqnarray}
{}^{(1)}J_{\mu}^{~\nu} =
-\frac{2}{3\ell^2} \left({}^{(1)}K_{\mu}^{~\nu} +2{}^{(1)}K \delta_{\mu}^{~\nu}\right)
\end{eqnarray}
and
\begin{eqnarray}
{}^{(1)}P_{\mu\alpha}^{~~~\nu\alpha}=-\left(
R_{\mu}^{~\nu} -\frac{1}{2}\delta_{\mu}^{~\nu} R 
\right)
\end{eqnarray}
into the junction condition, we can rewrite Eq.~(\ref{jnk1st}) as
%
\begin{eqnarray}
(1-\beta){}^{(1)}K_{\mu}^{~\nu}& = & -\frac{\kappa^2}{2}
\left(T_{\mu}^{~\nu} -\frac{1}{3}\delta_{\mu}^{~\nu} T \right)
\nonumber\\&&\quad
+\beta \ell \left(R_{\mu}^{~\nu} - \frac{1}{6} \delta_{\mu}^{~\nu}R\right). 
\label{jnk1st-2}
\end{eqnarray}
Then Eq.~(\ref{jnk1st-2}) together with Eq.~(\ref{1stsol}) yields the 
gravitational equations on the brane at first order:
%
\begin{equation}
R_{\mu}^{~\nu} -\frac{1}{2}\delta_{\mu}^{~\nu}R
 = \frac{\kappa^2}{\ell(1+ \beta)}T_{\mu}^{~\nu}
 +\frac{2}{\ell}\frac{1-\beta}{1+\beta} {}^{(1)} \chi_\mu^{~\nu}.
\label{ein_1st}
\end{equation}
%
We can see that
for vanishing $^{(1)} \chi_\mu^{~\nu}$
Einstein gravity is reproduced at this order.

From the coefficient of $T_{\mu}^{~\nu}$ we can read off the four-dimensional gravitational 
constant at low energies as
%
\begin{equation}
8\pi G= \frac{\kappa^2}{\ell (1+\beta)}. 
\label{g-beta}
\end{equation}
%
In the linear perturbation analysis of Ref.~\cite{DS}
the scale-dependent gravitational coupling was obtained,
and Eq.~(\ref{g-beta}) agrees with the result of ~\cite{DS}
at long distances.
Note, however, that the field equations~(\ref{ein_1st})
are nonlinear.
This indicates that nonlinearity of gravity 
does not affect the gravitational 
coupling at low energies/long distances. 

The integration constant
${}^{(1)}\chi_\mu^\nu$ is constrained from the Codacci equation:
%
\begin{eqnarray}
&&D_\nu \biggr[ {}^{(1)}K_{\mu}^{~\nu} -\delta_{\mu}^{~\nu} {}^{(1)}K
\nonumber\\
&&\quad
+2\alpha \left(3{}^{(1)}J_{\mu}^{~\nu} -\delta_{\mu}^{~\nu} {}^{(1)}J
+\frac{2}{\ell}{}^{(1)}P_{\mu}^{~\nu}\right) \biggr]=0,
\end{eqnarray}
which implies that
\begin{equation}
(1-\beta)  D_\nu \left[{}^{(1)}K_{\mu}^{~\nu} -\delta_{\mu}^{~\nu} {}^{(1)}K 
\right] =0. 
\end{equation}
%
Substituting Eq.~(\ref{1stsol}) into the above we obtain 
%
\begin{equation}
D_\nu {}^{(1)}\chi_{\mu}^{~\nu}=0. 
\end{equation}
%
Namely, ${}^{(1)}\chi_{\mu}^{~\nu}$ is transverse and traceless 
with respect to the brane geometry.
This means that it corresponds to the dark radiation, and
in fact we have $^{(1)}E_{\mu}^{~\nu}=-(2/\ell){}^{(1)}\chi_{\mu}^{~\nu}$ on the brane.
In what follows we set ${}^{(1)}\chi_{\mu}^{~\nu}=0$ for simplicity. 

\subsection{Second order equations}

We now go on to the second order calculations
in order to see corrections to
four-dimensional general relativity (other than $^{(1)}\chi_{\mu}^{~\nu}$).

First let us focus on the trace part.
The Hamiltonian constraint implies that 
%
\begin{eqnarray}
(1-\beta){}^{(2)}M & = & 
-\alpha \biggl[ \frac{1}{6}{}^{(1)}M^2-2 {}^{(1)}\tilde M_\mu^{~\nu} 
{}^{(1)}\tilde M_\nu^{~\mu} \nonumber \\
& & +{}^{(1)}L^{\alpha\beta}_{~~~\rho\sigma}
{}^{(1)}L_{\alpha\beta}^{~~~\rho\sigma} \biggr]. 
\end{eqnarray}
%
From the definition of $M$ and $L_{\alpha\beta}^{~~~\rho\sigma}$ we have
%
\begin{eqnarray}
{}^{(2)}M= \frac{6}{\ell}{}^{(2)}K -{}^{(1)}K^2+{}^{(1)}K_\mu^{~\nu} 
{}^{(1)}K_\nu^{~\mu}
\end{eqnarray}
%
and
%
\begin{eqnarray}
{}^{(1)}L_{\alpha\beta}^{~~~\rho\sigma}
=C_{\alpha\beta}^{~~~\rho\sigma},
\end{eqnarray}
%
where $C_{\alpha\beta\rho\sigma}$ is the four-dimensional 
Weyl tensor. Solving for ${}^{(2)}K$, we arrive at
%
\begin{eqnarray}
&&{}^{(2)}K  =   \frac{\ell}{6}\left[{}^{(1)}K^2-{}^{(1)}K_\mu^{~\nu} 
{}^{(1)}K_\nu^{~\mu}\right] \nonumber \\
& &\hspace{-5mm} -\frac{\beta}{1-\beta} \frac{\ell^3}{24} 
\left[ {\cal W}
-2{}^{(1)}\tilde M_\mu^{~\nu} {}^{(1)}\tilde M_\nu^{~\mu}
+\frac{1}{6}{}^{(1)}M^2 \right],
\end{eqnarray}
where ${\cal W}$ is the trace of
\begin{eqnarray}
{\cal W}_{\mu}^{~\nu}:=C_{\mu\alpha}^{~~~\beta\gamma}C_{\beta\gamma}^{~~~\nu\alpha}.
\end{eqnarray}
Recalling that ${}^{(1)}M=0$ and $^{(1)}\tilde M_\mu^{~\nu} =(2/\ell){}^{(1)}\chi_\mu^{~\nu}$
which is assumed to vanish, 
we obtain 
%
\begin{eqnarray}
{}^{(2)}K= -\frac{\ell^3}{24} \left(R_\mu^{~\nu} R_\nu^{~\mu} 
-\frac{1}{3}R^2 \right)
-\frac{\beta}{1-\beta} \frac{\ell^3}{24} {\cal W}.
\end{eqnarray}

Next let us consider the traceless part.
Again we need to solve the evolution equation
to obtain the corresponding part of the brane extrinsic curvature.

A simple calculation leads to
%
\begin{eqnarray}
{}^{(2)}_{(a)}\bar H_{\mu}^{~\nu} & = &  2\tilde{\cal W}_{\mu}^{~\nu}
-2{}^{(1)}\tilde M_\alpha^{~\beta} C_{\mu\beta}^{~~~\nu\alpha}
\nonumber\\
&&-2\left[{}^{(1)}\tilde M_{\mu}^{~\alpha} {}^{(1)}\tilde M_{\alpha}^{~\nu}
-\frac{1}{4}\delta_{\mu}^{~\nu}
{}^{(1)}\tilde M_{\alpha}^{~\beta} {}^{(1)}\tilde M_{\beta}^{~\alpha}
 \right] \nonumber \\
& & +\frac{2}{\ell^2}\frac{1+\beta}{1-\beta}{}^{(2)}\tilde M_{\mu}^{~\nu},
\\
{}^{(2)}_{(b)}\bar H_\mu^{~\nu} & = & 6
\left[{}^{(1)}\tilde M_{\mu}^{~\alpha} {}^{(1)}\tilde M_{\alpha}^{~\nu}
-\frac{1}{4}\delta_{\mu}^{~\nu}
{}^{(1)}\tilde M_{\alpha}^{~\beta} {}^{(1)}\tilde M_{\beta}^{~\alpha}
\right]
\nonumber \\
& &+6 {}^{(1)}\tilde M_\alpha^{~\beta} C_{\mu\beta}^{~~~\nu\alpha} 
-\frac{6}{\ell^2}{}^{(2)}E_\mu^{~\nu},
\\
{}^{(2)}_{(c)}\bar H_\mu^{~\nu}& =& 0,
\end{eqnarray}
%
where we defined the traceless part
$\tilde{\cal W}_{\mu}^{~\nu}$ of ${\cal W}_{\mu}^{~\nu}$.
Then the traceless part of the MT effective equations reduces to
%
\begin{eqnarray}
{}^{(2)}E_{\mu}^{~\nu} & = & -{}^{(2)}\tilde M_{\mu}^{~\nu} 
-\frac{\beta \ell^2}{3(1-\beta)}\frac{1}{a^4}\tilde {\cal W}_{\mu}^{~\nu}.
\label{e-m-2}
\end{eqnarray}
%
Note here that we have $1/a^4(y)$ factored out
so that $\tilde {\cal W}_{\mu}^{~\nu}$ is computed from
the induced metric $h_{\mu\nu}(x)$.
We again omitted the dark radiation term
${}^{(1)}\chi_{\mu}^{~\nu}\propto{}^{(1)}\tilde M_{\mu}^{~\nu}$.

From the definition of $\tilde M_{\mu}^{~\nu}$ we have
%
\begin{eqnarray}
{}^{(2)}\tilde M_{\mu}^{~\nu} & = & {}^{(2)}\tilde R_{\mu}^{~\nu} 
+{}^{(1)}K_\mu^{~\alpha} {}^{(1)}K_\alpha^{~\nu}
-\frac{1}{4}\delta_{\mu}^{~\nu}
{}^{(1)}K_\alpha^{~\beta} {}^{(1)}K_\beta^{~\alpha} \nonumber \\
& & - {}^{(1)} K {}^{(1)}\tilde K_{\mu}^{~\nu} 
+\frac{2}{\ell}{}^{(2)}\tilde K_{\mu}^{~\nu}.
\label{defM2}
\end{eqnarray}
%
Using Eqs.~(\ref{e-m-2}) and~(\ref{defM2}), we can write
the evolution equation as
%
\begin{eqnarray}
\mbox \pounds_n {}^{(2)}\tilde K_{\mu}^{~\nu}
& = & {}^{(2)}\tilde R_{\mu}^{~\nu}+\frac{4}{\ell}{}^{(2)}\tilde K_{\mu}^{~\nu}
-{}^{(1)}K^{(1)}\tilde K_{\mu}^{~\nu} \nonumber \\
& & +\frac{\beta \ell^2}{2(1-\beta)}\frac{1}{a^4}\tilde {\cal W}_\mu^{~\nu}.
\label{2ndevo}
\end{eqnarray}
The Ricci tensor $^{(2)}\tilde R_{\mu}^{~\nu}$
in Eq.~(\ref{2ndevo}) can be calculated as follows.
Integrating the first order result~(\ref{1stsol}),
we obtain the four-dimensional part of the bulk metric as
\begin{eqnarray}
&&\hspace{-5mm}q_{\mu\nu}(y,x) =a^2(y)
\left[ h_{\mu\nu}(x)+{}^{(1)}g_{\mu\nu}(y,x)+\cdots\right] \nonumber \\
& &= a^2h_{\mu\nu}+\frac{\ell^2}{2}\left(a^2-1\right)
\left[R_{\mu\nu}-\frac{1}{6}h_{\mu\nu}R \right] +\cdots,
\end{eqnarray}
where the boundary condition $^{(1)}g_{\mu\nu}(0,x)=0$ is understood.
Then the Ricci tensor is expressed as
\begin{eqnarray}
{}^{(2)}\tilde R_\mu^{~\nu}(y,x) 
& = & \frac{\ell^2}{2}\left(a^{-4}-a^{-2}\right) \biggl[ 
R_\mu^{~\alpha} R_\alpha^{~\nu} -\frac{1}{6}RR_{\mu}^{~\nu} \nonumber \\
& & -\frac{1}{4} \delta_{\mu}^{~\nu}\left(
R_\alpha^{~\beta} R_\beta^{~\alpha} -\frac{1}{6}
R^2 \right)\nonumber\\
&&-\frac{1}{2}\left(D^\alpha D_\nu R_{\alpha}^{~\mu}
+D_\alpha D^\mu R_\nu^{~\alpha}\right) \nonumber \\
& & +\frac{1}{3}{}D_\mu D^\nu R
+\frac{1}{2}D^2R_{\mu}^{~\nu} 
-\frac{1}{12}\delta_{\mu}^{~\nu}D^2 R
\biggr] \nonumber \\
&& \hspace{-8mm}=: \frac{\ell^2}{2}\left(a^{-4}-a^{-2}\right)
\left[{\cal S}_{\mu}^{~\nu} (x)
+\frac{1}{6}R \tilde R_{\mu}^{~\nu}\right],
\label{SR^2}
\end{eqnarray}
where $D^2:=D_{\alpha}D^{\alpha}$ and 
one should notice that ${\cal S}_{\mu}^{~\nu}$
satisfies ${\cal S}_{\mu}^{~\mu}=0$ and
\begin{eqnarray}
D_{\nu}{\cal S}_{\mu}^{~\nu}=0.
\end{eqnarray}
Using Eq.~(\ref{SR^2}) and the first order result
\begin{eqnarray}
^{(1)}K^{(1)}\tilde K_{\mu}^{~\nu}=\frac{\ell^2}{12}\frac{1}{a^4}R[h]\tilde R_{\mu}^{~\nu}[h],
\label{kkrr}
\end{eqnarray}
we can integrate
the evolution equation~(\ref{2ndevo}),
yielding
%
\begin{eqnarray}
{}^{(2)}\tilde K_{\mu}^{~\nu} & = & 
\frac{\ell^2}{2}\left(
\frac{y}{a^4}+\frac{\ell}{2}\frac{1}{a^2}  \right)
{\cal S}_{\mu}^{~\nu} +\frac{1}{24}\frac{\ell^3}{a^2}R \tilde R_{\mu}^{~\nu} \nonumber \\
& & +\frac{\beta \ell^2}{2(1-\beta)} \frac{y}{a^4}\tilde {\cal W}_{\mu}^{~\nu}
+\frac{{}^{(2)}\chi_{\mu}^{~\nu}(x)}{a^4},
\end{eqnarray}
where ${}^{(2)}\chi_{\mu}^{~\nu}$ is an integration constant
dependent only on the brane coordinates $x^{\mu}$.
To make the physical meaning of this integration constant clear,
we define
\begin{eqnarray}
{}^{(2)}\bar \chi_{\mu}^{~\nu}(x)
& := & {}^{(2)}\chi_\mu^\nu+\frac{\ell^3}{4}{\cal S}_\mu^\nu 
+\frac{\beta \ell^3}{6(1-\beta)} {\tilde{\cal W}}_\mu^\nu \nonumber \\
& & \quad+\frac{\ell^3}{8} \Biggl[R_\mu^{~\alpha}R_\alpha^{~\nu}
-\frac{1}{3}RR_{\mu}^{~\nu} \nonumber \\
& & \qquad-\frac{1}{4}\delta_{\mu}^{~\nu} \left( 
R_\alpha^{~\beta}R_\beta^{~\alpha}-\frac{1}{3}R^2 \right) \Biggr].
\end{eqnarray}
Now we can see
that $^{(2)}E_{\mu}^{~\nu}=-(2/\ell) {}^{(2)}\bar \chi_{\mu}^{~\nu}$ on the brane.
Thus we finally have the extrinsic curvature of the brane
in terms of $\bar\chi_{\mu}^{~\nu}$:
\begin{eqnarray}
^{(2)}K_{\mu}^{~\nu}(y=0, x)&=&
\frac{\ell^3}{24}\biggl[ -3R_\mu^{~\alpha}R_\alpha^{~\nu}
+2R R_{\mu}^{~\nu} \nonumber \\
& & \hspace{-15mm}
-\frac{5}{12}\delta_{\mu}^{~\nu} R^2  
+ \frac{1}{2}\delta_{\mu}^{~\nu}R_{\alpha}^{~\beta}R_{\beta}^{~\alpha}
\biggr] \nonumber \\
&&\hspace{-20mm}
-\frac{\beta \ell^3}{6(1-\beta)}\left[{\cal W}_\mu^{~\nu}
-\frac{3}{16}\delta_{\mu}^{~\nu}{\cal W}
\right]+{}^{(2)}\bar \chi_{\mu}^{~\nu},
\end{eqnarray}
which can be rearranged into
\begin{eqnarray}
&&{}^{(2)}K_{\mu}^{~\nu}-\delta_{\mu}^{~\nu}{}^{(2)}K
=\frac{\ell^3}{2}{\cal P}_{\mu}^{~\nu}
\nonumber \\
&&\quad-\frac{\beta\ell^3}{6(1-\beta)}\left[{\cal W}_{\mu}^{~\nu} 
-\frac{7}{16}\delta_{\mu}^{~\nu}
{\cal W}\right]+{}^{(2)}\bar\chi_{\mu}^{~\nu}, \label{sol}
\end{eqnarray}
where
\begin{eqnarray}
{\cal P}_{\mu}^{~\nu}&:=&
\frac{1}{6}RR_{\mu}^{~\nu}-\frac{1}{4}R_{\mu}^{~\alpha}R_{\alpha}^{~\nu}
\nonumber\\&&\qquad
+\frac{1}{8}\delta_{\mu}^{~\nu}R_{\alpha}^{~\beta}R_{\beta}^{~\alpha}
-\frac{1}{16}\delta_{\mu}^{~\nu}R^2.
\end{eqnarray}
The form of ${\cal P}_{\mu}^{~\nu}$ is suggestive.
Rewriting this in terms of a new variable defined by
$s_{\mu}^{~\nu}:=R_{\mu}^{~\nu}-(1/2)\delta_{\mu}^{~\nu}R$,
we obtain the same expression as the well-known
quadratic energy-momentum term which was
first introduced in Ref.~\cite{SMS}.

The Codacci equation at second order is summarized as
%
\begin{eqnarray}
D_{\nu}\left[(1-\beta)\left({}^{(2)}K_{\mu}^{~\nu}-\delta_{\mu}^{~\nu}{}^{(2)}K\right)
+\frac{\beta \ell^3}{2}{\cal Y}_{\mu}^{~\nu}
\right]=0,
\end{eqnarray}
%
where
%
\begin{eqnarray}
{\cal Y}_{\mu}^{~\nu} & := &  R_{\mu\alpha}^{~~~\nu\beta} R_{\beta}^{~\alpha}
- R  R_{\mu}^{~\nu}+\frac{3}{2} R_{\mu}^{~\alpha}R_{\alpha}^{~\nu} \nonumber \\
& & -\frac{3}{4}\delta_{\mu}^{~\nu}R_{\alpha}^{~\beta}R_{\beta}^{~\alpha}
+\frac{7}{24}\delta_{\mu}^{~\nu} R^2
\nonumber \\
&=&C_{\mu\alpha}^{~~~\nu\beta} R_{\beta}^{~\alpha}
-2{\cal P}_{\mu}^{~\nu}.
\end{eqnarray}
The Codacci equation is formally integrated to give 
%
\begin{eqnarray}
& & (1-\beta)\left({}^{(2)}K_{\mu}^{~\nu}-\delta_{\mu}^{~\nu}{}^{(2)}K\right)
+\frac{\beta \ell^3}{2}{\cal Y}_{\mu}^{~\nu} \nonumber \\
& &\qquad= \tau_\mu^{~\nu} (x) + c_1 {\cal S}_\mu^{~\nu}  + c_2 
{\cal Z}_\mu^{~\nu}
\label{localandnonlocal}
\end{eqnarray}
%
where $\tau_{\mu}^{~\nu}$ is the part that
cannot be expressed in terms of local quantities.
As for the local part,
${\cal Z}_{\mu}^{~\nu}$ is the divergence free tensor defined by 
%
\begin{eqnarray}
{\cal Z}_{\mu}^{~\nu}:=  R  R_{\mu}^{~\nu} -\frac{1}{4}\delta_{\mu}^{~\nu} 
R^2 - D_\mu  D^\nu  R +\delta_{\mu}^{~\nu}  D^2  R,
\end{eqnarray}
%
and $c_1$ and $c_2$ are constant coefficients.
Note that ${\cal S}_{\mu}^{~\nu}$ and ${\cal Z}_{\mu}^{~\nu}$
are the two linearly independent,
divergence free combinations of curvature tensors
which are of order $\epsilon^2$.
This is because the variation of
$\sqrt{-h}R^2$ and $\sqrt{-h}R_{\mu\nu}R^{\mu\nu}$
with respect to the metric gives ${\cal Z}_{\mu}^{~\nu}$
and ${\cal S}_{\mu}^{~\nu}+{\cal Z}_{\mu}^{~\nu}/3$, respectively.
Due to the Gauss-Bonnet theorem in four dimensions,
another curvature squared term
$\sqrt{-h}R_{\mu\nu\rho\sigma}R^{\mu\nu\rho\sigma}$
does not give rise to a new independent local tensor quantity.

Substituting Eq.~(\ref{sol}) into Eq.~(\ref{localandnonlocal}), we obtain
\begin{eqnarray}
\tau_{\mu}^{~\nu}(x) &=& \frac{\beta \ell^3}{2}{\cal Y}_{\mu}^{~\nu}
-c_1{\cal S}_{\mu}^{~\nu} -c_2{\cal Z}_{\mu}^{~\nu}
\nonumber\\
&&\quad+(1-\beta) \Biggl[{}^{(2)}\bar \chi_\mu^{~\nu}
+\frac{\ell^3}{2}{\cal P}_{\mu}^{~\nu}
\nonumber\\
&&\qquad
-\frac{\beta \ell^3}{6(1-\beta)} \left({\cal W}_\mu^{~\nu}
-\frac{7}{16}\delta_\mu^{~\nu} {\cal  W} \right)
\Biggr].
\end{eqnarray}
This equation relates the integration constant $^{(2)}\bar\chi_{\mu}^{~\nu}$
to the nonlocal part $\tau_{\mu}^{~\nu}$ and the free parameters $c_1$ and $c_2$.
The traceless condition $^{(2)}\bar \chi_\mu^{~\mu}=0$ yields
the constraint
%
\begin{eqnarray}
\tau_\mu^{~\mu} & = & (1-3\beta ) \frac{\ell^3}{8}
\left[ R_\mu^{~\nu}  R_\nu^{~\mu} -\frac{1}{3} R^2 \right] \nonumber \\
& &\qquad+\frac{\beta \ell^3}{8}{\cal W}-3c_2D^2 R.
\end{eqnarray}
%
In the cases of $\beta=0$ (the Randall-Sundrum model),
$\tau_\mu^{~\mu}$ represents the 
trace anomaly, which is expected from the AdS/CFT correspondence~\cite{adscft} 
in the braneworld model~\cite{adscftbrane}.
The nonlocal nature of $\tau_{\mu\nu}$ indicates that 
it corresponds to the energy-momentum of the holographic CFT on the brane. 
However, 
it is not trivial whether this is the case when $\beta\neq0$,
because the
AdS/CFT correspondence in the presence of the Gauss-Bonnet term
has not been addressed much yet~\cite{adscftGBbrane}.

Including the second order computation
in the junction condition,
we shall derive the 
effective equations on the brane
with corrections to four-dimensional general relativity. 
The junction condition now becomes 
%
\begin{eqnarray}
& & (1-\beta )\left({}^{(1)}K_\mu^{~\nu} -\delta_\mu^{~\nu} {}^{(1)}K 
+ {}^{(2)}K_{\mu}^{~\nu}-\delta_{\mu}^{~\nu}{}^{(2)}K \right) \nonumber \\
& & =-\frac{\kappa^2}{2}T_{\mu}^{~\nu}-\frac{\beta \ell^3}{2}{\cal Y}_{\mu}^{~\nu}
+ \beta \ell  G_\mu^{~\nu} .
\end{eqnarray}
%
Using the results obtained in the previous and present subsections,
we have
\begin{eqnarray}
G_\mu^{~\nu} =\frac{\kappa^2 }{\ell(1+\beta)}T_\mu^{~\nu} 
+\frac{2}{\ell(1+\beta)}  \left( \tau_\mu^{~\nu} + c_1{\cal S}_\mu^{~\nu} 
+c_2 {\cal Z}_\mu^{~\nu}\right), \nonumber\\\label{effective1}
\end{eqnarray}
or
\begin{eqnarray}
G_\mu^{~\nu}&=&\frac{\kappa^2 }{\ell(1+\beta)}T_\mu^{~\nu}
+\frac{2}{\ell}\frac{1-\beta}{1+\beta}{}^{(2)}\bar\chi_\mu^{~\nu}
\nonumber\\
&&\quad
+\frac{(1-3\beta)\ell^2}{1+\beta}{\cal P}_\mu^{~\nu}
+\frac{\beta\ell^2}{1+\beta}C_{\mu\alpha}^{~~~\nu\beta}R_{\beta}^{~\alpha}
\nonumber\\
&&\qquad
-\frac{\beta\ell^2}{3}
\left[{\cal W}_\mu^{~\nu}
-\frac{7}{16}\delta_\mu^{~\nu} {\cal  W} \right].
\label{effective2}
\end{eqnarray}
Here the constants
$c_1$ and $c_2$ are determined by the boundary condition
other than that imposed at the brane. 
We stress that
the above equations are correct even for nonlinear gravity, as long as 
they are applied to the low energy regime or at long distances.

\section{Application: Cosmology on the brane}

As an application, let us consider homogeneous and isotropic cosmology on the brane.
We write the induced metric as
\begin{eqnarray}
h_{\mu\nu}dx^{\mu}dx^{\nu}=-dt^2+a^2(t)\delta_{ij}dx^idx^j,
\end{eqnarray}
where $a(t)$ is the scale factor. 
For this metric the Weyl tensor vanishes, $C_{\mu\nu\rho\sigma}=0$,
and hence we have
${\cal W}_{\mu}^{~\nu}=0$ and ${\cal Y}_{\mu}^{~\nu}=-2{\cal P}_{\mu}^{~\nu}$.
Non-vanishing components of ${\cal P}_{\mu}^{~\nu}$ are given by
\begin{eqnarray}
{\cal P}_{t}^{~t}&=&-\frac{3}{4}H^4,
\\
{\cal P}_{i}^{~j}&=&-\left(\frac{3}{4}H^4+H^2\dot H\right)\delta_{i}^{~j},
\end{eqnarray}
where $H:=\dot a/a$ and a dot stands for a derivative with respect to $t$.
Then a straightforward calculation shows that
\begin{eqnarray}
D_{\nu}{\cal Y}_{\mu}^{~\nu}=-2D_{\nu}{\cal P}_{\mu}^{~\nu}=0,
\end{eqnarray}
and thus the Codacci equation at
second order implies 
\begin{eqnarray}
D_{\nu}{}^{(2)}\bar\chi_{\mu}^{~\nu}=0.
\end{eqnarray}
This means that ${}^{(2)}\bar\chi_{\mu}^{~\nu}$
behaves as a conserved radiation like component on the cosmological brane
(as is the case at first order).
Assuming that this dark radiation term vanishes, we obtain
the modified Friedmann equation
from the $(t t)$ component of the effective equations~(\ref{effective2}):
\begin{eqnarray}
H^2=\frac{\kappa^2}{3\ell(1+\beta)}\rho+\frac{(1-3\beta)\ell^2}{4(1+\beta)}H^4.
\label{feq:ge}
\end{eqnarray}

An exact form of the modified Friedmann equation
in the Gauss-Bonnet braneworld was derived in
Refs.~\cite{Charm, MT}, which,
omitting the dark radiation, is summarized as
\begin{eqnarray}
\kappa^2(\rho+\sigma)=2\sqrt{H^2+\ell^{-2}}
\left(3-\beta+2\beta\ell^2 H^2\right).
\label{feq.full}
\end{eqnarray}
Expanding the right hand side of Eq.~(\ref{feq.full})
for small $\epsilon=\ell^2 H^2$, we obtain the same expression as Eq.~(\ref{feq:ge}).
Thus the validity of the gradient expansion is confirmed.

As long as one considers the homogeneous and isotropic universe 
on the brane, there is less advantage of our effective equations
compared to \cite{MT}. 
However, it is sure that our procedure has a great advantage for general cases 
without symmetry. This is because we have the low energy 
effective theory at the nonlinear level.

\section{Summary and Discussion}

In this paper we have obtained
the effective gravitational equations
at low energies in the ``Gauss-Bonnet'' braneworld.
The derivation here is
along the geometrical projection approach of
Maeda and Torii~\cite{MT}.
At low energies we can solve iteratively
the evolution equation in the bulk
by expanding the relevant equations
in terms of $\epsilon:=(\ell/L)^2 \ll 1$,
where $\ell$ is the bulk curvature scale
and $L$ is the brane intrinsic curvature scale.
Up to second order in the gradient expansion,
nonlinear gravity on the Gauss-Bonnet brane
is described by Eq.~(\ref{effective1}) or Eq.~(\ref{effective2}),
which is the main result of the present paper.
Although
the Gauss-Bonnet term makes
the original governing equations complicated and far from transparent,
our effective equations have
the form of the Einstein equations with correction terms,
and are simple enough to handle. 
We can repeat the same procedure to include higher order effects.

In the low energy effective equations~(\ref{effective1})
we have
the nonlocal tensor $\tau_{\mu\nu}$ and
two free parameters corresponding to the bulk degrees of freedom.
In the Randall-Sundrum model ($\beta=0$),
we can determine that part to close the equations
via the AdS/CFT correspondence
once we specify a CFT on the brane~\cite{KS}.
However, such a holographic interpretation of the bulk geometry is obscure
in the presence of the Gauss-Bonnet term,
and this point is an open question for further study.
Not relying on the holographic argument, we can instead
determine the integration constant $\chi_{\mu\nu}$ by imposing
another boundary condition, for example, at a second brane
introduced away from the first brane.


\section*{Acknowledgements}

TS thanks Kei-ichi Maeda for useful discussion in the early stage 
of this paper. 
TK is supported by the JSPS under Contract No.~01642.
The work of TS was supported by Grant-in-Aid for Scientific 
Research from Ministry of Education, Science, Sports and Culture of 
Japan (No.~13135208, No.~14102004, No.~17740136 and No.~17340075), 
the Japan-U.K. and Japan-France Research  Cooperative Program.



\appendix

\section{Comment on an alternative formulation}

The main text is based on the Gauss-Bonnet brane equations derived by Maeda and Torii~\cite{MT} 
using a $(4+1)$ decomposition first developed by Shiromizu \textit{et al.}~\cite{SMS} to derive the 
Einstein brane equations. In this appendix we sketch a similar formulation of Gauss-Bonnet brane 
equations developed in Ref.~\cite{DK} in the case of Einstein gravity,
with an emphasis on the structure of the equations.

Consider a five dimensional spacetime in gaussian normal coordinates $x^A=(x^\rho,y)$ with 
line element 
\begin{eqnarray}
ds^2= dy^2+q_{\mu\nu}(x^\rho,y)dx^\mu dx^\nu \label{aeq1}
\end{eqnarray}
and expand the metric coefficients $q_{\mu\nu}(x^\rho,y)$ near the surface $y=0$ as
\begin{eqnarray}
q_{\mu\nu}(x^\rho,y) & = & g_{\mu\nu}(x^\rho)+k_{\mu\nu}(x^\rho)\,y+{1\over2}l_{\mu\nu}(x^\rho)\,y^2 
\nonumber \\
& & +{\cal O}(y^3)\,.\label{aeq2} 
\end{eqnarray}
At lowest order in $y$
the Riemann tensor of the metric~(\ref{aeq1}) is
\begin{eqnarray}
& & {\cal R}_{y\mu y\nu}=-{1\over2}l_{\mu\nu}+{1\over4}k_{\rho\mu}k^\rho_{~\nu},
\nonumber \\
& & {\cal R}_{y\mu \nu\rho}=-{1\over2}D_\nu k_{\mu\rho}+{1\over2}D_\rho k_{\mu\nu},
\label{aeq3} \\
& & {\cal R}_{\mu \nu\rho\sigma}=R_{\mu \nu\rho\sigma} +{1\over4}(k_{\mu\sigma}k_{\nu\rho}
-k_{\mu\rho}k_{\nu\sigma}), \nonumber 
\end{eqnarray}
where $D_\mu$ and $R_{\mu \nu\rho\sigma}$ are 
the covariant derivative and Riemann tensor of the metric $g_{\mu\nu}(x^\rho)$. 

The Einstein-Gauss-Bonnet field equations are
\begin{eqnarray}
\Lambda \delta^{~A}_B+{\cal G}^{~A}_B+\alpha\,{\cal H}^{~A}_{B}=0 \label{aeq4}
\end{eqnarray}
where ${\cal G}_B^{~A}:= {\cal R}_B^{~A}-(1/2)\delta_B^{~A}{\cal R}$ is 
the five-dimensional
Einstein tensor and where the Gauss-Bonnet tensor ${\cal H}_B^{~A}$ is defined as
\begin{eqnarray}
{\cal H}_B^{~A} & := & 2 \left[{\cal R}^{ALMN}{\cal R}_{BLMN}-2{\cal R}^{LM}{\cal R}^A_{~~LBM}
\right.\nonumber \\
& & \left.-2 {\cal R}^{AL}{\cal R}_{BL}+{\cal R}{\cal R}_B^{~A} \right]
-{1\over2}\delta_B^{~A}{\cal L}_{\rm GB}.  
\label{aeq5}
\end{eqnarray}

Expanding the Einstein tensor at lowest order in $y$ is a straightforward calculation
which yields
\begin{eqnarray}
{\cal G}_{yy} & = &
-\frac{1}{8}\left(g_{\mu\beta}g_{\nu\alpha}-g_{\mu\nu}g_{\alpha\beta}\right)k^{\mu\nu}
k^{\alpha\beta}-{1\over2}R,
\\
{\cal G}_{y\mu} & = & {1\over2}\left(g_{\mu\beta}g_{\nu\alpha}-g_{\mu\nu}g_{\alpha\beta}\right)
D^\nu k^{\alpha\beta}
\nonumber\\
&=&\frac{1}{2}D_{\alpha}\left(k^\alpha_{~\mu}-\delta^\alpha_{~\mu}k\right),
 \label{aeq6} \\
{\cal G}_{\mu\nu} & = &
-\frac{1}{2}\left(g_{\mu\beta}g_{\nu\alpha}-g_{\mu\nu}g_{\alpha\beta}\right)l^{\alpha\beta}
+\frac{1}{2}k^{\alpha}_{~\mu}k_{\alpha\nu}-\frac{1}{4}kk_{\mu\nu}
\nonumber\\&&\quad
+{1\over8}g_{\mu\nu}(k^2-3k^\rho_{~\sigma} k^\sigma_{~\rho})
+G_{\mu\nu}.  
\end{eqnarray} 
Expanding the Gauss-Bonnet tensor is only slightly more
involved and yields
\begin{eqnarray}
{\cal H}_{yy}&=&
-{1\over16}\left(
N_{\mu\alpha\nu\beta}-4P_{\mu\alpha\nu\beta}\right)k^{\mu\nu}k^{\alpha\beta}
\cr&&\quad
-{1\over2}\left(
R_{\mu\alpha\nu\beta}R^{\mu\alpha\nu\beta}-4R_{\alpha\beta}R^{\alpha\beta}+R^2\right),
\\
{\cal H}_{y\mu}&=&
{1\over2}N_{\mu\alpha\nu\beta}D^\nu k^{\alpha\beta}
\nonumber\\&=&
{1\over2}D_\alpha\left({3\over2}j^{~\alpha}_\mu
-{1\over2}\delta^{~\alpha}_\mu j-4P^\alpha_{\ \beta\mu\gamma}k^{\beta\gamma}\right),
\label{eqno(7)}\\
{\cal H}_{\mu\nu}&=&-{1\over2}N_{\mu\alpha\nu\beta}l^{\alpha\beta}
+2\overline{\cal M}_{\mu\nu}-{1\over2}g_{\mu\nu}\overline{{\cal L}},
\end{eqnarray}
where $P^{\mu\nu\rho\sigma}$ is defined in Eq.~(\ref{def=P}) in the main text,
and the fully developed expressions of the other tensors are
\begin{widetext}
\begin{eqnarray}
j^{\mu\nu}&:=&
-{2\over3}k^{\mu\rho}k_{\rho\sigma}k^{\sigma\nu}
+{2\over3}k\,k^{\mu\rho}k_{~\rho}^\nu+{1\over3}k^{\mu\nu}
\left(k^\rho_{~\sigma} k^\sigma_{~\rho}-k^2\right),
\\
N^{\mu\alpha\nu\beta}&:=&
-\left(
k^{\mu\beta}k^{\alpha\nu}-k^{\mu\nu}k^{\alpha\beta}
\right)
+{1\over2}\left(
g^{\mu\beta}g^{\alpha\nu}-g^{\mu\nu}g^{\alpha\beta}
\right)(k^\rho_{~\sigma} k^\sigma_{~\rho}-k^2)
-k^{\mu\rho}k_{~\rho}^\beta g^{\alpha\nu}+k^{\alpha\rho}
k_{~\rho}^\beta g^{\mu\nu}
\cr&&\quad
-k^{\alpha\rho}k^\nu_{~\rho}
g^{\mu\beta}+k^{\mu\rho}k^\nu_{~\rho} g^{\alpha\beta}
-kk^{\alpha\beta}g^{\mu\nu}
-kk^{\mu\nu}g^{\alpha\beta}
+kk^{\mu\alpha}g^{\nu\beta}+kk^{\nu\beta}g^{\mu\alpha}-
4P^{\mu\alpha\nu\beta},
\label{eqno(8)}\\
\overline{\cal M}_{\mu\nu}&:=&
-{3\over4}k_{~\mu}^\alpha k_{~\nu}^\beta k_{~\alpha}^\gamma k_{\beta\gamma}
+{3\over8}k_{~\rho}^{\sigma}k_{~\sigma}^{\rho}k_{~\mu}^\alpha k_{\nu\alpha}
+{1\over4}k_{~\rho}^{\sigma}k_{~\lambda}^{\rho}k^{\lambda}_{~\sigma}k_{\mu\nu}
+{5\over8}kk_{~\mu}^\alpha k_{~\nu}^\beta k_{\alpha\beta}
-{5\over16}k k_{~\rho}^{\sigma}k_{~\sigma}^{\rho} k_{\mu\nu}
\cr&&\quad
-{1\over4}k^2k_{~\mu}^\alpha k_{\nu\alpha}
+{1\over16}k^2k_{\mu\nu}
+{1\over2}\left(R_{\mu\alpha\beta\gamma}k_{~\nu}^\gamma k^{\alpha\beta}
+R_{\nu\alpha\beta\gamma}k_{~\mu}^\gamma k^{\alpha\beta}\right)
-{1\over2}R_{\alpha\beta}k_{~\mu}^\alpha k_{~\nu}^\beta
+{1\over2}R_{\alpha\beta}k^{\alpha\beta}k_{\mu\nu}
\cr&&\qquad
-{1\over4}R_{\mu\alpha\nu\beta}k^{\alpha\gamma}k^\beta_{~\gamma}
+{1\over2}kR_{\mu\alpha\nu\beta}k^{\alpha\beta}
-k^{\alpha\beta}\left(R_{\mu\alpha}k_{\nu\beta}
+R_{\nu\alpha}k_{\mu\beta}\right)
+{1\over2}k\left(R_{\mu\alpha}k^\alpha_{~\nu}+R_{\nu\alpha}k^\alpha_{~\mu}\right)
\cr&&\quad\qquad
+{3\over4}R_{\mu\nu}(3k_{~\rho}^{\sigma}k_{~\sigma}^{\rho}-k^2)
+{1\over2}R\left(2k_{~\mu}^\alpha k^\alpha_{~\nu}-kk_{\mu\nu}\right),
\\
\overline{{\cal L}}&:=&
-{7\over8}k^{\alpha}_{~\beta}k^{\beta}_{~\gamma}k^{\gamma}_{~\lambda}k^{\lambda}_{~\alpha}
+{7\over16}\left(k_{~\rho}^{\sigma}k_{~\sigma}^{\rho}\right)^2
+kk^{\alpha}_{~\beta} k^{\beta}_{~\gamma} k^{\gamma}_{~\alpha}
-{5\over8}k^2(k_{~\rho}^{\sigma}k_{~\sigma}^{\rho})+{1\over16}k^4
\cr&&\quad
+R_{\alpha\beta\gamma\delta}k^{\alpha\delta}k^{\beta\gamma}
-4R_{\alpha\beta}k^{\alpha\gamma}k^\beta_{~\gamma}
+2kk^{\alpha\beta}R_{\alpha\beta}
+{3\over2}Rk_{~\rho}^{\sigma}k_{~\sigma}^{\rho}
-{1\over2}Rk^2.
\end{eqnarray}
\end{widetext}
There are no quadratic terms such as $R_{\mu\alpha}R^{\alpha}_{~\nu}$
in ${\cal H}_{\mu\nu}$ as the Gauss-Bonnet tensor is identically zero in four dimensions.
As is well-known~\cite{Nat_rev}, neither ${\cal H}_{yy}$
nor ${\cal H}_{y\mu}$ contain $l_{\mu\nu}$;
as for ${\cal H}_{\mu\nu}$ it is quasi-linear in $l_{\mu\nu}$.


The Gauss-Bonnet field equations at lowest order in $y$
can be seen as a boundary condition problem:
given a $4$-dimensional metric $g_{\mu\nu}$ and an extrinsic curvature $k_{\mu\nu}/2$
satisfying the constraints [the $(yy)$ and $(y\mu)$ components of the field equations],
then the $(\mu\nu)$ component gives $l_{\mu\nu}$ which is the metric in the bulk at order
${\cal O}(y^2)$.
When $\alpha\neq0$, $l_{\mu\nu}$ is only known implicitly because of the quasi-linearity of the Gauss-Bonnet tensor ${\cal H}_{AB}$.

Now the junction conditions relate the extrinsic curvature to the energy-momentum tensor
on the brane~\cite{Israel_Junk, Davis:2002gn}.
They can in fact be read off from the $(y\mu)$ component of the
field equations,
and are written in terms of $k_{\mu\nu}$ and $j_{\mu\nu}$ as
\begin{eqnarray}
T^\mu_{~\nu}&=&
\sigma\delta^\mu_{~\nu}-k^\mu_{~\nu}+\delta^\mu_{~\nu} k
\cr&&\quad
-\alpha\left({3\over2}j^\mu_{~\nu}-{1\over2}\delta^\mu_{~\nu}
j-4P^\mu_{\ \rho\nu\sigma}k^{\rho\sigma}\right), \label{eqno(10)}
\end{eqnarray}
where the constant $\sigma:=(6/\ell)\left[1-(4\alpha/3\ell^2)\right]$
is such that $k_{\mu\nu}=-(2/\ell)g_{\mu\nu}$, together with $R_{\mu\nu\rho\sigma}=0$,
solves the brane equations~(\ref{aeq4}),
$\ell$ being related to $\Lambda$ by
$\Lambda=-(6/\ell^2)\left(1-2\alpha/\ell^2\right)$.
(Here and hereafter we set $\kappa^2=1$.)
The $(y\mu)$ component of the brane field equations is then
equivalent to the conservation of the energy-momentum tensor: $D_{\nu}T_{\mu}^{~\nu}=0$.

In Einstein gravity ($\alpha=0$),
inverting the junction equations to express $k_{\mu\nu}$
in terms of $T_{\mu\nu}$ is elementary:
$k_{\mu\nu}=-(2/\ell)g_{\mu\nu}-\left[T_{\mu\nu}-(1/3)Tg_{\mu\nu}\right]$. 
Hence the brane field equations (\ref{aeq4}) can readily be written in terms of $T_{\mu\nu}$ rather than 
$k_{\mu\nu}$, and we arrive at 
the effective equations on the brane~\cite{SMS}
\begin{eqnarray}
G_{\mu\nu}& = & \kappa_4^2 T_{\mu\nu}- E_{\mu\nu}-{1\over4}T_{\mu\rho}T^\rho_{~\nu}
+{1\over12}TT_{\mu\nu} \cr
& & \quad+{1\over8}g_{\mu\nu}T^\rho_{~\sigma}
 T^\sigma_{~\rho}-{1\over24}T^2g_{\mu\nu},\label{aeq12}
\end{eqnarray}
with the identification $\kappa_4^2=1/\ell$.
In deriving the above equation we introduced the electric part of the Weyl tensor
and rewrote $l_{\mu\nu}$ as
\begin{eqnarray}
l_{\mu\nu}=-2E_{\mu\nu}-\frac{1}{3}\Lambda g_{\mu\nu}+\frac{1}{2}k_{\mu\rho}k^{\rho}_{~\nu}.
\end{eqnarray}

In Einstein-Gauss-Bonnet gravity,
the junction equations~(\ref{eqno(10)}) 
cannot explicitly be inverted to give $k_{\mu\nu}$ in terms of $T_{\mu\nu}$. 
Hence, the gradient expansion scheme developed in the main text,
which consists in writing $k_{\mu\nu}=-(2/\ell)g_{\mu\nu}+\epsilon_{\mu\nu}$
and expanding~(\ref{eqno(10)}) to second order in $\epsilon_{\mu\nu}$,
is a way
to solve iteratively the brane equations in a closed form.





\begin{thebibliography}{99}

\bibitem{Maa_rev}
  D.~Langlois,
  Prog.\ Theor.\ Phys.\ Suppl.\  {\bf 148}, 181 (2003)
  [arXiv:hep-th/0209261],
  R.~Maartens,
  Living Rev.\ Rel.\  {\bf 7}, 7 (2004)
  [arXiv:gr-qc/0312059].
  P.~Brax, C.~van de Bruck and A.~C.~Davis,
  Rept.\ Prog.\ Phys.\  {\bf 67}, 2183 (2004)
  [arXiv:hep-th/0404011],
  C.~Csaki,
  arXiv:hep-ph/0404096.

\bibitem{RS}
L.~Randall and R.~Sundrum,
  Phys.\ Rev.\ Lett.\  {\bf 83}, 4690 (1999)
  [arXiv:hep-th/9906064],
L.~Randall and R.~Sundrum,
  Phys.\ Rev.\ Lett.\  {\bf 83}, 3370 (1999)
  [arXiv:hep-ph/9905221].

\bibitem{Nat_rev}
  N.~Deruelle and J.~Madore,
  arXiv:gr-qc/0305004.


\bibitem{Charm}
  C.~Charmousis and J.~F.~Dufaux,
  Class.\ Quant.\ Grav.\  {\bf 19}, 4671 (2002)
  [arXiv:hep-th/0202107].


\bibitem{Cos_others}
  N.~Deruelle and T.~Dolezel,
  Phys. Rev. {\bf D62}, 103502 (2000)
  [arXiv:gr-qc/0004021],
  S.~Nojiri, S.~D.~Odintsov and S.~Ogushi, 
  Phys. Rev. {\bf D65},  023521(2002) 
  [arXiv:hep-th/0108172],
  C.~Germani, C.~F.~Sopuerta, 
  Phys. Rev. Lett. {\bf 88}, 231101 (2002)
  [arXiv:hep-th/0202060],
  J.~E.~Lidsey, S.~Nojiri and S.~D.~Odintsov,
  JHEP {\bf 0206}, 026 (2002)
  [arXiv:hep-th/0202198],
  J.~E.~Lidsey and N.~J.~Nunes,
  Phys.\ Rev.\ D {\bf 67}, 103510 (2003)
  [arXiv:astro-ph/0303168],
  J.~F.~Dufaux, J.~E.~Lidsey, R.~Maartens and M.~Sami,
  Phys.\ Rev.\ D {\bf 70}, 083525 (2004)
  [arXiv:hep-th/0404161],
  S.~Tsujikawa, M.~Sami and R.~Maartens,
  Phys.\ Rev.\ D {\bf 70}, 063525 (2004)
  [arXiv:astro-ph/0406078],
  K.~s.~Aoyanagi and K.~i.~Maeda,
  Phys.\ Rev.\ D {\bf 70}, 123506 (2004)
  [arXiv:hep-th/0408008],
  M.~Sami, N.~Savchenko and A.~Toporensky,
  Phys.\ Rev.\ D {\bf 70}, 123528 (2004)
  [arXiv:hep-th/0408140],
  T.~Kobayashi,
  Gen.\ Rel.\ Grav.\  {\bf 37}, 1869 (2005)
  [arXiv:gr-qc/0504027].
  

\bibitem{DS}
  N.~Deruelle and M.~Sasaki,
  Prog.\ Theor.\ Phys.\  {\bf 110}, 441 (2003)
  [arXiv:gr-qc/0306032].

\bibitem{others}
  J.~E.~Kim, B.~Kyae and H.~M.~Lee,
  Nucl.\ Phys.\ B {\bf 582}, 296 (2000)
  [Erratum-ibid.\ B {\bf 591}, 587 (2000)]
  [arXiv:hep-th/0004005],
   J.~E.~Kim and H.~M.~ Lee, 
  Nucl. Phys. {\bf B602}, 346(2001)
  [arXiv:hep-th/0010093],
  K.~A.~Meissner and M.~Olechowski,
  Phys.\ Rev.\ D {\bf 65}, 064017 (2002)
  [arXiv:hep-th/0106203],
  I.~P.~Neupane,
  Phys.\ Lett.\ B {\bf 512}, 137 (2001)
  [arXiv:hep-th/0104226],
  I.~P.~Neupane,
  Class.\ Quant.\ Grav.\  {\bf 19}, 5507 (2002)
  [arXiv:hep-th/0106100],
  Y.~M.~Cho and I.~P.~Neupane,
  Int.\ J.\ Mod.\ Phys.\ A {\bf 18}, 2703 (2003)
  [arXiv:hep-th/0112227],
  M.~Minamitsuji and M.~Sasaki,
  Prog.\ Theor.\ Phys.\  {\bf 112}, 451 (2004)
  [arXiv:hep-th/0404166].

\bibitem{MT}
K.~i.~Maeda and T.~Torii,
  Phys.\ Rev.\ D {\bf 69}, 024002 (2004)
  [arXiv:hep-th/0309152].

\bibitem{SMS}
T.~Shiromizu, K.~i.~Maeda and M.~Sasaki,
  Phys.\ Rev.\ D {\bf 62}, 024012 (2000)
  [arXiv:gr-qc/9910076].

\bibitem{BT}
G.~W.~Gibbons and S.~W.~Hawking,
  Phys.\ Rev.\ D {\bf 15}, 2752 (1977),
R.~C.~Myers,
  Phys.\ Rev.\ D {\bf 36}, 392 (1987).

\bibitem{DK}
  N.~Deruelle and J.~Katz,
  Phys.\ Rev.\ D {\bf 64}, 083515 (2001)
  [arXiv:gr-qc/0104007].


\bibitem{Israel_Junk}
 W.~Israel,
  Nuovo Cim.\ B {\bf 44S10}, 1 (1966)
  [Erratum-ibid.\ B {\bf 48}, 463 (1967\ NUCIA,B44,1.1966)].

\bibitem{Davis:2002gn}
  S.~C.~Davis,
  Phys.\ Rev.\ D {\bf 67}, 024030 (2003)
  [arXiv:hep-th/0208205],
  E.~Gravanis and S.~Willison,
  Phys.\ Lett.\ B {\bf 562}, 118 (2003)
  [arXiv:hep-th/0209076].

\bibitem{KS}
S.~Kanno and J.~Soda,
  Phys.\ Rev.\ D {\bf 66}, 043526 (2002)
  [arXiv:hep-th/0205188].

\bibitem{GE}
T.~Wiseman,
  Class.\ Quant.\ Grav.\  {\bf 19}, 3083 (2002)
  [arXiv:hep-th/0201127],
T.~Shiromizu and K.~Koyama,
  Phys.\ Rev.\ D {\bf 67}, 084022 (2003)
  [arXiv:hep-th/0210066],
S.~Kanno and J.~Soda,
  Gen.\ Rel.\ Grav.\  {\bf 36}, 689 (2004)
  [arXiv:hep-th/0303203].


\bibitem{adscft}
J.~M.~Maldacena,
  Adv.\ Theor.\ Math.\ Phys.\  {\bf 2}, 231 (1998)
  [Int.\ J.\ Theor.\ Phys.\  {\bf 38}, 1113 (1999)]
  [arXiv:hep-th/9711200].

\bibitem{adscftbrane}
S.~S.~Gubser,
  Phys.\ Rev.\ D {\bf 63}, 084017 (2001)
  [arXiv:hep-th/9912001],
S.~B.~Giddings, E.~Katz and L.~Randall,
  JHEP {\bf 0003}, 023 (2000)
  [arXiv:hep-th/0002091],
T.~Shiromizu and D.~Ida,
  Phys.\ Rev.\ D {\bf 64}, 044015 (2001)
  [arXiv:hep-th/0102035],
S.~Nojiri, S.~D.~Odintsov and S.~Zerbini,
  Phys.\ Rev.\ D {\bf 62}, 064006 (2000)
  [arXiv:hep-th/0001192],
S.~Nojiri and S.~D.~Odintsov,
  Phys.\ Lett.\ B {\bf 484}, 119 (2000)
  [arXiv:hep-th/0004097],
L.~Anchordoqui, C.~Nunez and K.~Olsen,
  JHEP {\bf 0010}, 050 (2000)
  [arXiv:hep-th/0007064],
S.~W.~Hawking, T.~Hertog and H.~S.~Reall,
  Phys.\ Rev.\ D {\bf 62}, 043501 (2000)
  [arXiv:hep-th/0003052],
K.~Koyama and J.~Soda,
  JHEP {\bf 0105}, 027 (2001)
  [arXiv:hep-th/0101164],
A.~Padilla,
  Class.\ Quant.\ Grav.\  {\bf 23}, 3983 (2006)
  [arXiv:hep-th/0512083].

\bibitem{adscftGBbrane}
J.~P.~Gregory and A.~Padilla,
  Class.\ Quant.\ Grav.\  {\bf 20}, 4221 (2003)
  [arXiv:hep-th/0304250],
S.~Ogushi and M.~Sasaki,
  Prog.\ Theor.\ Phys.\  {\bf 113}, 979 (2005)
  [arXiv:hep-th/0407083].


\end{thebibliography}
\end{document}